\documentclass[a4paper]{article}
\usepackage{color}
\usepackage{amsthm}
\usepackage{amsmath}
\usepackage{graphicx}
\usepackage{amssymb}
\usepackage{bbold}
\usepackage{esint}
\usepackage{subfigure}
\usepackage{color}
\usepackage[all]{xy}

\usepackage{anysize}
\marginsize{2cm}{2cm}{2cm}{2cm}

\def\beq{\begin{equation}}
\def\eeq{\end{equation}}
\def\bea{\begin{eqnarray}}
\def\eea{\end{eqnarray}}

\makeatletter

\theoremstyle{plain}

  \theoremstyle{remark}

\begin{document}

\title{Local realizations of anyon exchange symmetries without lattice dislocations}
%\begin{center}
\author{Miguel Jorge Bernabe Ferreira,$^{a}$\footnote{migueljb@if.usp.br} ~Pramod Padmanabhan,$^{a,b}$\footnote{pramod23phys@gmail.com}~\\ Paulo Teotonio-Sobrinho$^{a}$\footnote{teotonio@fma.if.usp.br}  } 
%\end{center}
\maketitle

\begin{center}
{\small $^{a}$ Departamento de F\'{i}sica Matem\'{a}tica, Universidade de S\~{a}o Paulo}

{\small  Rua do Mat\~ao Travessa R Nr.187, S\~ao Paulo, CEP 05508-090}

{\small $^{b}$ Department of Physics and Astronomy, University of California, Irvine}

{\small 4129 Frederick Reines Hall Irvine, CA 92697-4575}
\end{center}

\begin{abstract}

The global $e$-$m$ exchange symmetry of the toric code is realized locally through an exactly solvable Hamiltonian on a two dimensional lattice which has no lattice dislocations and their associated defect line. The Hamiltonian is still changed locally in selected sites where we wish to realize this anyon symmetry. We refer to these selected sites as defect sites in analogy with the usual lattice defects. The operators on the defect sites condense dyons of the toric code and are shown to support states which obey the fusion rules of Ising anyons just as the lattice dislocations thus achieving the transition to the Ising phase from the toric code phase. They can also be introduced in an entire region leading to an idea of non-localized defects. The method leads to a natural generalization for other Abelian groups where they help realize all the anyon exchange symmetries locally.

\end{abstract}

\section*{Introduction}
~

Physics in three dimensional space is governed by bosons and fermions as the fundamental group is the permutation group, $S_n$ \cite{BalBook} which have two irreducible one dimensional representations. The story changes in two dimensions where the fundamental group is given by the braid group, $B_n$ of $n$ particles which leads to a rich variety of statistics. These are termed as anyons \cite{Wilc} which have recently found applications in the field of topological quantum computation \cite{freed, nayak}. Exactly solvable lattice models have been used to obtain these particles as low energy excitations of which the most famous one is the toric code model of Kitaev and it's variants \cite{kitToric, aguado1}. These are examples of systems with long-ranged entangled ground states in which information in qubits could be encoded non-locally. Soon they were realized to be a major part of the classification of phases of matter at low temperatures and were defined to possess a new type of order called topological order \cite{h1}.

The phase of the toric code model is described by it's anyon content which are the representations of the quantum double of $\mathbb{Z}_2$ \cite{kitToric}. They are the electric charge, $e$, magnetic flux, $m$, the combination of the two called the dyon, $\epsilon$ and the vacuum, 1 which can be thought of as a trivial anyon. They obey the fusion rules 
$ e\times m = \epsilon ~;~ e\times e = m\times m =\epsilon \times \epsilon = 1~;~e\times \epsilon =m~;~m\times\epsilon =e $
with the other fusion rules being the trivial ones. The dyon is a real fermion and the rest are bosons. It is easy to see that these rules are invariant under an interchange of $e$ and $m$ labels. This is the electric-magnetic duality of the toric code model which was known earlier \cite{sd}. In recent literature this global exchange symmetry was seen to be part of the anyon symmetry group \cite{wang, frad}. Gauging this symmetry leads to new topologically ordered phases from a given parent phase with interesting features such as the emergence of non-Abelian anyons from Abelian phases. Recently efforts have been made to construct models realizing these symmetries locally with the earliest examples being the work of Bombin \cite{Bombin} where a microscopic model with lattice defects were introduced to gauge the electric-magnetic symmetry of the $\mathbb{Z}_2$ toric code. This process led to the emergence of Ising anyons in a model where the underlying degrees of freedom belonged to that of the toric code. Thus this served as an example of realizing non-Abelian anyons from an Abelian system \cite{pp2}. Although Abelian systems are simpler to be experimentally implemented, they are not enough for quantum computation \cite{Pachos}. Finding Abelian systems supporting non-Abelian anyons not only deepens our understanding of topological phases but is also relevant for quantum computation \cite{bravyi, walker, bond}. This model was further generalized for the $\mathbb{Z}_n$ case in \cite{wen3}. Ising anyons were shown to be obtainable from the toric code anyons as a fusion category by superposing toric code anyons in \cite{Pachos, Pachos2}. These features were also studied for the toric code in \cite{kk}. These efforts were taken further in a more complete theory of such defects in Ablelian theories in the effective field theory language in \cite{maisham} where such symmetries were also realized in bi-layer fractional quantum Hall states \cite{maisham2}. 
A mathematical theory was developed recently under the name of $G$-crossed braided fusion categories in \cite{wang}. Recently global realizations of these symmetries were obtained as models for symmetry enriched topological (SET) phases in \cite{herm, fid}. Despite these important advances there are still very few exactly solvable microscopic lattice models that realize these phases locally in a systematic manner.

In this article we propose a microscopic, exactly solvable two dimensional lattice model which just uses the familiar operators appearing in the description of the $\mathbb{Z}_2$ toric code to realize the electric-magnetic symmetry locally.  Moreover we also show that Ising anyons emerge in this phase. We do not introduce any physical defect/dislocation in the lattice to achieve this. However we change the Hamiltonian at sites, $s=(v,p)$ where the electric-magnetic symmetry is realized locally. We call these sites as defect sites and they should not be confused with the lattice defects used in \cite{Bombin}. The operators associated with these defect sites were introduced earlier in \cite{ppm}. These defect sites act as sources and sinks for the dyons. The eigenvalues of the operators associated to the defect sites define two sectors just as in the case of the defects of \cite{Bombin}. The difference in this model is that the defect site achieves both the exchange of $e$ and $m$ and when we have two such defect sites we can realize the Ising anyon phase as well. This is in contrast to the lattice defects introduced in \cite{Bombin} where the defect operators at the ends of the branch cut are very important for realizing the Ising anyon phase. Another important difference is that we can now create {\it defect regions} which consists of an area of defect sites. This helps switch the charge and flux in a non-local way. When we have two such defect regions then we can fuse them in the same way as the localized defect sites to again realize the Ising anyon phase. By comparing the degeneracies of this model with those using lattice dislocations \cite{wen3} we obtain a relation between the two approaches. Finally the model presented here can be easily generalized to other Abelian cases and possibly non-Abelian ones as well. In particular the model naturally realizes all the anyon exchange symmetries of the $\mathbb{Z}_N$ toric code anyons.

\section*{The model}

The setup of the model is similar to that of the toric code with the degrees of freedom located on the links of a two dimensional lattice. For simplicity we choose this to be a square lattice though the construction goes through for a lattice with an arbitrary triangulation just like the toric code model. 

%The local Hilbert space, $\mathcal{H}_l$ is generated by the following basis $\{\vert +1\rangle,\vert -1\rangle\}$ with the total Hilbert space given by $\bigotimes_l \mathcal{H}_l$. 

%A state of the system will be a linear combination of vectors of the following form
%$$\vert l_1, l_2,\cdots,l_{n_l}\rangle=\vert l_1\rangle\otimes\vert l_2\rangle\otimes\cdots\otimes \vert l_{n_l}\rangle\;,$$
%where $n_l$ is the number of links of the lattice.

\begin{figure}[h!]
\begin{center}
		\includegraphics[scale=1]{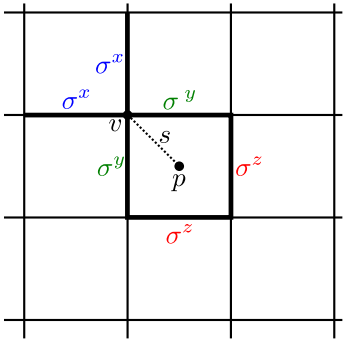}
	\caption{Action of the operator $O_s$ on a site $s=(v,p)$. The $\sigma^x$ operator acts on the {\it spikes} of the vertex $v$. The remaining Pauli matrices act on the sides of the plaquette $p$.}
\label{action}
\end{center}
\end{figure}

The dynamics is given by vertex $A_v$ and plaquette $B_p$ operators, defined as the usual toric code stabilizers
\beq A_v =\bigotimes_{l\in\hbox{star}(v)}\sigma_l^x \;\;\;\;\; \hbox{and} \;\;\;\;\; B_p=\bigotimes_{l\in\partial p}\sigma_l^z\;. \eeq
Apart from these operators we also consider operators that act on the sites of the lattice, called site operators $O_s$. Each site $s=(v,p)$ is formed by a vertex and a plaquette as shown in figure \ref{action}. The site operator is define as
$$O_{s^\prime}=A_{v^\prime}B_{p^\prime}\;,\;\; \hbox{where} \;\; s^\prime=(v^\prime,p^\prime)\;.$$
The action of this operator on the qubits is shown in figure \ref{action}.

Consider $S=\{s_0,s_1,\cdots,s_n\},$ a set of sites of the lattice. The Hamiltonian is given by
\beq
\label{H}
H=-\sum_{s\in S}O_s-\sum_{p\notin S}B_p - \sum_{v\notin S}A_v\;,
\eeq
where by $p\notin S$ ($v\notin S$) we mean the plaquette $p$ (vertex $v$) is not part of any site which belongs to $S$. It is easy to check that this model is exactly solvable and it's spectrum can be readily written down just as in the toric code case. The ground states satisfy
$B_p\vert \Psi^0\rangle=A_v\vert \Psi^0\rangle=O_s\vert \Psi^0\rangle=\vert \Psi^0\rangle.$
Note that the ground states of the toric code are also ground states of this model. The difference arises due to $O_s\vert \Psi^0 \rangle=\vert \Psi^0 \rangle$, which can also be written as $A_vB_p\vert \Psi^0 \rangle=\vert \Psi^0 \rangle$ ($s=(v,p)$) which implies $A_v\vert \Psi^0 \rangle=B_p\vert \Psi^0 \rangle=\vert \Psi^0 \rangle$ or $A_v\vert \Psi^0 \rangle=B_p\vert \Psi^0 \rangle=-\vert \Psi^0 \rangle.$
This implies that the dyon on site $s\in S$ gets condensed into the ground state. On a closed manifold the toric code excitations can only be created in pairs, thus if a given state $\vert \Psi\rangle$ has an even number of dyonic excitations localized on the sites $s\in S$ (and no other kind of excitation anywhere else) it is still in the vacuum of the Hamiltonian \eqref{H}. The ground state degeneracy of this model is given by the sum of all possible ways of creating a pair of dyons on the $n$ sites of $S$ and is given by $2^{2g+n-1}.$ In the case of the torus this reduces to $2^{n+1}$ for a system with $n$ defect site operators. In earlier realizations of the electric-magnetic duality using lattice dislocations the ground state degeneracy was found to be $2^{\frac{m}{2}+1}$ for $m$ lattice dislocations \cite{wen3} and this also implies that the dislocations carry non-integral quantum dimension hinting at their non-Abelian behavior. Comparing the two ground state degeneracies we see that $n$ defect site operators correspond to $2n$ lattice dislocations which implies that each defect site operator mimics the behavior of two lattice dislocations. Since the defect sites are a combination of two lattice dislocations, we may ask how the phase with non-Abelian behavior arises in this model. We shall see that the model supports states of the form $e+m$ apart from the usual toric code anyons which behave like Ising anyons as shown in \cite{Pachos}.  

\section*{Excitations}
~

The excitations of this model are similar to the excitations of the toric code except in the region where we have the defect sites that is the region where the Hamiltonian is given by the operators $O_{s_i}$ acting on the sites $s_i$. In the rest of the lattice the excitations are the deconfined charges, $e$, fluxes, $m$ and the dyons, $\epsilon$. In the Hamiltonian in \eqref{H} we have $n$ sites $s_i$ where the operator $O_{s_i}$ acts. We will consider different cases starting with the case where we have a single isolated defect site, that is $n=1$. Then we will consider the case with $n>1$ acting on isolated sites. Finally we will consider the case where we still have $n>1$, but now the defect sites are not isolated but grouped into a region covering an area $A$. In what follows we will only see the behavior of the excitations near the defect site or the defect region.  

\section*{Single defect site ($n=1$)}
~

The site operator $O_{s}$ has two eigenvalues, $\pm 1$ with the corresponding states denoted by $\sigma_\pm$. They split the Hilbert space into two sectors with the excitations of the system behaving differently depending on which sector the state is in.

\begin{figure}[h!]
\centering
\subfigure[]{
\includegraphics[scale=1]{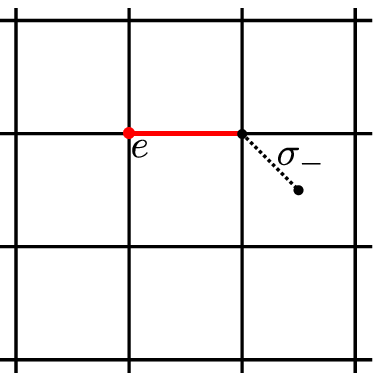}\label{singlee}
}
\hspace{1cm}
\subfigure[]{
\includegraphics[scale=1]{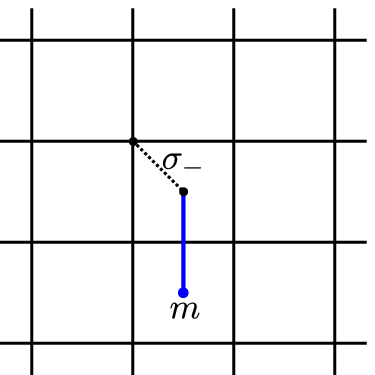}\label{singlem}
}
\caption{An isolated charge or flux for which the $\sigma_-$ sector is a source or sink of. The red strings correspond to string operators formed out of $\sigma^z$ operators and the blue strings correspond to string operators made of $\sigma^x$ operators. }
\label{seila}
\end{figure}

Consider the creation of a sector where the eigenvalue of the site operator is $-1$ or the $\sigma_-$ state. A way of doing this is by creating an isolated charge near the defect site which also excites the $O_s$ operator into the $\sigma_-$ sector as shown in figure \ref{singlee}.
A similar situation occurs for the flux excitations of the system as shown in figure \ref{singlem}. The $\sigma_-$ sector acts as a sink or source for the charge (and also the flux) as shown in figures \ref{singlee} and \ref{singlem} respectively. Thus in this sector the system can support a state of the form $e+m$ that is an isolated charge and a flux \cite{Pachos}.

\begin{figure}[h!]
\centering
\subfigure[]{
\includegraphics[scale=1]{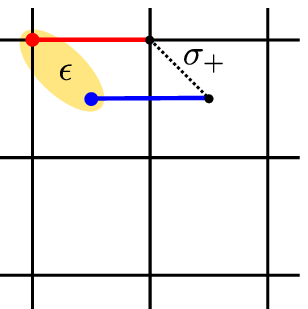}\label{mtoe2}
}
\hspace{.01cm}
\subfigure[]{
\includegraphics[scale=1]{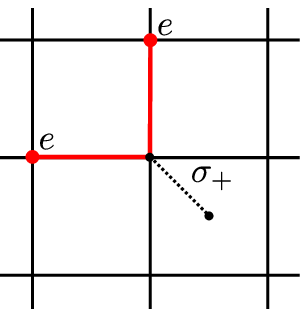}\label{paire}
}
\hspace{.01cm}
\subfigure[]{
\includegraphics[scale=1]{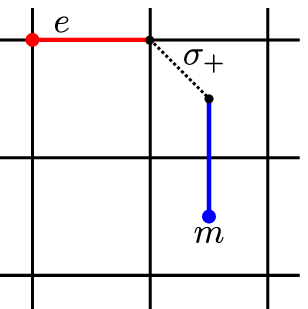}\label{mtoe}
}
\caption{The anyons that can occur in the $\sigma_+$ sector, (a) an isolated dyon and (b) a pair of charges, and (c) the permutation of the $e$-$m$ excitations.}
\label{seila2}
\end{figure}

The $\sigma_+$ sector acts as source/sink for the dyonic excitations as shown in figure \ref{mtoe2}. 
Apart from acting as a source/sink for the dyonic excitation, the $\sigma_+$ sector can also host a pair of charges as shown in figure \ref{paire} or a pair of fluxes which are part of the deconfined excitations of the toric code. Thus in this sector the system can host a state of the form $1+\epsilon$. 

Another interesting feature of the $\sigma_+$ sector is that it can interchange a charge and a flux, as shown in figure \ref{mtoe}. This can be seen in the following way. The charge and the flux excitation that compose the dyonic excitation shown in figure \ref{mtoe2} can be moved apart from each other giving rise to the state shown in figure \ref{mtoe}. This implies that in this sector of the Hilbert space a charge excitation can be made into a flux excitation by crossing the defect site. Note that this occurs without the introduction of any lattice dislocation.

\section*{Isolated defect sites ($n>1$)}
~

As in the case with $n=1$ the defect sites act as sources and sinks for the toric code excitations according to the eigenvalues of the operators defined on those defect sites and as before each of them also switch a charge into a flux and vice versa. However the phase in which this model exists is different as the ground degeneracy is different from that of the toric code case unlike the case with a single defect site which has the same degeneracy as the toric code. The only new feature now is that we can ``fuse'' two defect sites by fusing the particles for which these defect sites are sources.

\begin{figure}[h!]
\centering
\subfigure[]{
\includegraphics[scale=1]{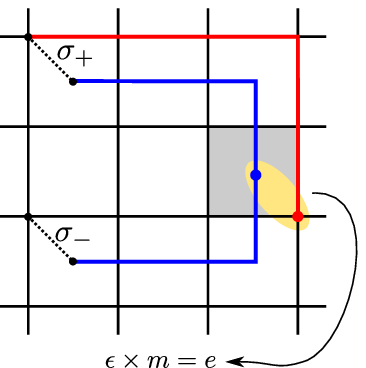}\label{fusionpme}
}
\hspace{.5cm}
\subfigure[]{
\includegraphics[scale=1]{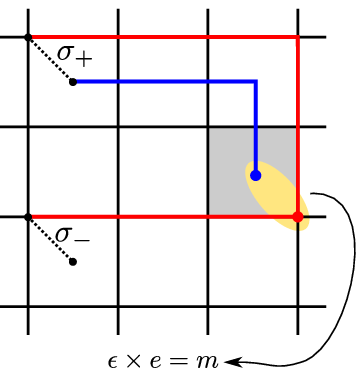}\label{fusionpmm}
}
\hspace{.5cm}
\subfigure[]{
\includegraphics[scale=1]{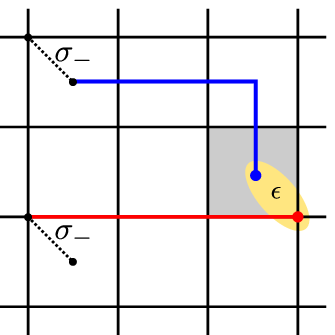}\label{fusionmme}
}
\hspace{.5cm}
\subfigure[]{
\includegraphics[scale=1]{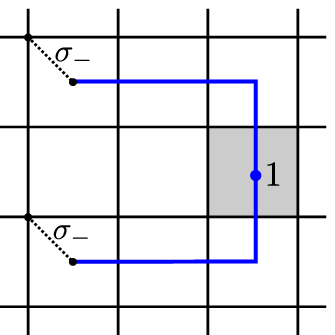}\label{fusionmm1}
}
\caption{Fusion rules of the $\sigma_+$ and $\sigma_-$ sectors, with the fusion channels of $\sigma_+$ and $\sigma_-$ giving (a) charge and (b) flux, and the fusion channels of $\sigma_-$ and $\sigma_-$ giving (c) a dyon and (d) the vacuum. }
\label{fusion}
\end{figure}

The way to fuse the two sectors $\sigma_\pm$ is by moving the anyons for which these sectors are a source of into a common site of the lattice. We can call these anyons as defect excitations. To move a defect excitation one has to extend the string that ends at the defect site. Figure \ref{fusionpme} shows a possible way of fusing $\sigma_+$ and $\sigma_-$ where the respective anyons are moved into the shaded region.
In the same way the $\sigma_-$ excitation could have also moved into the shaded region by applying a blue string as shown in figure \ref{fusionpmm}, in this case fusion would give $\epsilon\times m = e$, a charge excitation as the result. In a similar way we could have fused the vacuum in the $\sigma_+$ sectors with the $e$ or $m$ particles from the $\sigma_-$ sector resulting in a $e$ and $m$ particle respectively. Thus there are several channels for the fusion of the $\sigma_+$ and $\sigma_-$sectors leading to the non-Abelian fusion rule
$$\sigma_{\pm}\times \sigma_{\mp} = e+m\;.$$
Using the same procedure it is not difficult to see that the fusion rules for $\sigma_\pm\times \sigma_\pm$ are also non-Abelian and are given by
$$\sigma_\pm\times\sigma_\pm=1+\epsilon\;.$$
Figures \ref{fusionmme} and \ref{fusionmm1} illustrate these two ways for the fusion of two $\sigma_-$. This comes from the fact that there are two ways to move $\sigma_-$ (and also $\sigma_+$) excitations since they are source/sink for two kind of particles as shown in figure \ref{seila} and \ref{mtoe2}.

We can also fuse the anyons created far away from the defect site with the defects $\sigma_+$ and $\sigma_-$. By using the fusion rules of the toric code we easily obtain
$$ \sigma_{\pm}\times \epsilon = \sigma_{\pm}~;~\sigma_{\pm}\times e= \sigma_{\pm}\times m = \sigma_{\mp}.$$

%These are the fusion rules which gives rise to Ising anyons as noted in \cite{Bombin}. 
In particular we can choose $\sigma=e+m$ and $\psi=\epsilon$. Along with the vacuum 1 they give rise to the fusion rules of Ising anyons \cite{Bombin}.

It is easy to generalize this statement for a system with $n$ defect sites. The two sectors are given by the product of the eigenvalues of the operators $O_{s_1},\cdots , O_{s_n}$ on the defect sites, $s_1, \cdots , s_n$.

\section*{Defect sites in a region $A$}
~

Finally we consider the situation where the defect sites are grouped into a region $A$ with $n$ sites. This model has the same ground state degeneracy as the case where the defect sites are isolated. The features of the defect sites are unchanged from the previous situations including the realization of the fusion rules of the Ising anyons. The two sectors in this case are given by the possible product of the eigenvalues of the operators $O_{s_i}, i\in A$ which are $\pm1$ as before. We can think of this whole region as a defect which is spread in a region with area $A$. Thus we have a non-localized defect unlike the previous two situations where we dealt with isolated, localized defect sites. Inside this region there is no distinction between $e$ and $m$ excitations.

\begin{figure}[h!]
\begin{center}
		\includegraphics[scale=1]{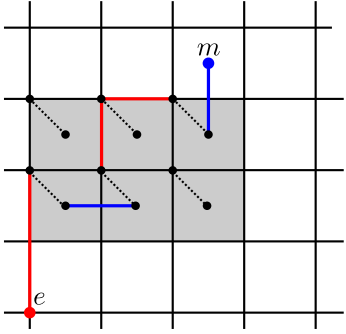}
	\caption{Non-local transmutation of a charge, $e$ and flux, $m$ in a defect region $A$. The area $A$ is represented by the shaded region.}
\label{e2mA}
\end{center}
\end{figure}

The new feature in this case which was not present in the earlier cases is that the transmutation of the charge and the flux can happen {\it non-locally}. By this we mean that a charge which enters a defect region $A$ can exit it from any point of the region as a flux, $m$ as shown in figure \ref{e2mA}. That is there exists a string operator which helps us achieve this. Figure \ref{e2mA} also shows an example of such a string operator. 

As before the two sectors are labeled by $\sigma_+$ and $\sigma_-$. If we have two isolated defect regions we can think of fusing them as we did in the case of isolated defects. We again have Ising anyons in this case as the argument for seeing this to be the case is similar to the previous situation. 

Finally we can think of the extreme case where we only have defect site operators all over the lattice. Then we are completely out of the Ising phase and are in a phase where there is no distinction between the charges and fluxes and all the dyons are condensed.

\section*{Other $\mathbb{Z}_N$ groups}

An advantage of this approach is that they can be naturally generalized to the case of other groups and inputs, which the parent quantum double models are based on. In particular they can be easily written down for the $\mathbb{Z}_N$ case where we have $N-1$ $\mathbb{Z}_2$ anyon exchange symmetries. The GSD of these models is given by $N^{2g + \sum_{i=1}^{N-1} n_i - 1}$ where $n_i$ denotes the number of defect sites which realize the $i$th anyon exchange symmetry. On the torus this reduces to the result in \cite{wen3}, $N^{n+1}$, where $n$ is a pair of dislocations realizing one of the anyon exchange symmetries. Each dislocation displays non-Abelian behavior while the defect site operators support superpositions of $Z_N$ toric code anyons which have non-Abelian behavior.

\section*{Remarks}

Decorating topological phases with dislocations were studied from the categorical point of view in \cite{kk} and their appearance can be understood as follows. Physical systems with boundaries have two phases, a bulk and a boundary phase. Depending on the type of the boundary, smooth or rough, we have different boundary phases. Domain walls between different topologically ordered phases contain a phase different from the phases on either side of it as the bulk anyons from the two phases fuse into the domain wall making them domain wall excitations. This problem reduces to the problem of identifying the boundary phase once we fold the system along the domain wall \cite{kk}. The notion of fusion of two such domain walls in the presence of three different topologically ordered phases into a single wall can then be introduced. Along similar lines joining two domain walls, between different topologically ordered phases, at a point results in a new phase locally at these points. These points have been studied as defect points or dislocations in the lattice. As mentioned before these are physical artifacts of the lattice on which the system lives on resulting in a change in the Hamiltonian locally around the regions where they are present. Thus the introduction of these lattice dislocations changed the Hamiltonian locally which described a new phase, in this case the Ising phase emerged from the toric code phase. The operators introduced in this paper achieve the effects of these defects/dislocations without a modification of the lattice through a new microscopic Hamiltonian which again described the Ising phase by condensing the dyons of the toric code phase.

\section{Acknowledgments}
~

The authors would like to thank FAPESP for support during this work. We are grateful to Hector Bombin for useful comments.


\begin{thebibliography}{99}

\bibitem{BalBook} A. P. Balachandran, G. Marmo, B. S. Skagerstam, A. Stern, {\it Classical Topology and Quantum States}, World Scientific, (1991).

\bibitem{Wilc} F. Wilczek, {\it Fractional Statistics and Anyon Superconductivity}, World Scientific, (1990).

\bibitem{h1} X. G. Wen, {\it  Quantum Field Theory of Many-body Systems: From the Origin of Sound to an Origin of Light and Electrons}, Oxford Graduate Texts, (2007).

\bibitem{freed} M.~H.~Freedman, A.~Kitaev, M.~J.~Larsen, Z.~Wang, Bull. Amer. Math. Soc. 40 (2003), 31-38. 

\bibitem{nayak} C.~Nayak, S.~H.~Simon, A.~Stern, M.~Freedman, S.~D.~Sarma, Rev.~Mod.~Phys.~80, 1083 (2008).

\bibitem{kitToric} A.~Y.~Kitaev, Ann.~Phys.~303, 2 (2003).

\bibitem{aguado1} O.~Buerschaper, J.~M.~Mombelli, M.~Christandl, M.~Aguado, J.~Math.~Phys.~54, 012201 (2013). 

\bibitem{sd} A. Kapustin, E. Witten, arXiv:hep-th/0604151.

\bibitem{wang} M.~Barkeshli, P.~Bonderson, M.~Cheng, Z.~Wang, arXiv:1410.4540 [cond-mat.str-el].

\bibitem{frad} J.~C.~Y.~Teo, T.~L.~Hughes, E.~Fradkin, Annals of Physics 360, 349 (2015). 

\bibitem{Bombin} H. Bombin, Phys.Rev.Lett.105:030403, (2010).

\bibitem{pp2} P. Padmanabhan, P. Teotonio-Sobrinho, Annals of Physics 361, pp. 266-277 (2015).

\bibitem{wen3} Yi-.~Z.~You, X-.~G.~Wen, Phys. Rev. B 86, 161107(R) (2012).

\bibitem{bravyi} S. Bravyi, Physical Review A 73, 42313 (2006).

\bibitem{walker}  M. Freedman, C. Nayak, K. Walker, Phys. Rev. B 73, 245307 (2006).

\bibitem{bond}  P. Bonderson, S. Sarma, M. Freedman, C. Nayak, arXiv:1003.2856 (2010).

\bibitem{Pachos} J.~R.~Wootton, V.~Lahtinen, Z.~Wang, J.~K.~Pachos, Phys. Rev. B 78, 161102(R) (2008).

\bibitem{Pachos2} J.~R.~Wootton, V.~Lahtinen, B.~Doucot, J.~K.~Pachos, Ann. Phys. 326, 2307 (2011).

\bibitem{kk} A.~Kitaev, L.~Kong, Commun. Math. Phys. 313 (2012) 351-373.

\bibitem{maisham} M. Barkeshli, C-M. Jian, X- L. Qi, Phys. Rev. B 88, 235103 (2013). 

\bibitem{maisham2} M. Barkeshli, X.-L. Qi, Phys. Rev. X 2, 031013 (2012).

\bibitem{herm} M. Hermele, Phys. Rev. B 90, 184418 (2014).

\bibitem{fid} N. Tarantino, N. Lindner, L. Fidkowski, arXiv:1506.06754 [cond-mat.str-el]. 

\bibitem{ppm} M. J. B. Ferreira,  P. Padmanabhan, P. Teotonio-Sobrinho, J. Phys. A: Math. Theor. 47 (2014) 375204 (50pp).

\end{thebibliography}
\end{document}